\documentclass[twocolumn,a4paper]{article} 
\expandafter\let\csname equation*\endcsname\relax
\expandafter\let\csname endequation*\endcsname\relax
\usepackage{siunitx,graphicx,overpic,epstopdf,amsmath,amssymb,bm}
\usepackage{tikz,tkz-tab}
\usepackage{subcaption}
\usepackage{cite}
\usepackage{gensymb}
\usepackage{authblk}

\begin{document}

\title{Analysis of Fourier ptychographic microscopy with half of the captured images}

\author[1,2]{Ao Zhou}
\author[1,*]{Ni Chen}
\author[1,2]{Haichao Wang}
\author[1]{Guohai Situ}
\affil[1]{Shanghai Institute of Optics and Fine Mechanics, Chinese Academy of Sciences, Shanghai 201800, China}
\affil[2]{University of Chinese Academy of Sciences, Beijing 100049, China.}

\affil[*]{Corresponding authors: nichen@siom.ac.cn}

\maketitle

\begin{abstract}
Fourier ptychography microscopy~(FPM) is a new computational imaging technique that can provide gigapixel images with both high resolution and a wide field of view~(FOV). However, time consuming of the data-acquisition process is a critical issue. In this paper, we make an analysis on the FPM imaging system with half number of the captured images. Based on the image analysis of the conventional FPM system, we then compare the reconstructed images with different number of captured data. Simulation and experiment results show that the reconstructed image with half number captured data do not show obvious resolution degradation compared to that with all the captured data, except a contrast reduction. In particular in the case when the object is close to phase-only/amplitude-only, the quality of the reconstructed image with half of the captured data is nearly as good as the one reconstructed with full data.
\end{abstract}

\section{Introduction}
It's well known that in the conventional microscope, the low numerical aperture~(NA) of the objective lens produces a wide FOV image, but induces low image resolution. FPM is a newly developed computational optical imaging technique, which breaks the diffraction limit of the objective lens by using an angularly varying light emitting diode~(LED) illumination and a phase retrieval algorithm~\cite{zheng2013wide,ou2013quantitative,zheng2015fourier}. In the FPM system, a programmable LED array is usually used as the light source. After capturing a stack of low resolution images under different illumination angles, an iterative phase retrieval process~\cite{Fienup2013,Chen_2014_JOSK} is used to reconstruct the object's complex field with enhanced resolution without sacrificing the wide FOV. Because the FPM breaks the limitation of the space-bandwidth product~(SBP) of the optical system, and achieves gigapixel imaging, it has a great potential in a variety of applications, such as biomedical imaging~\cite{dong2014sparsely,tian20143d,williams2014fourier,chung2015counting,tian2015computational,horstmeyer2015digital}, characterizing unknown optical aberrations of lenses~\cite{bian2013adaptive,ou2014embedded}.

However, FPM's high SBP imaging capability is time consuming, limiting its application to only static objects. This limitation comes from the requirement of a large amount of low resolution images. In addition, the low illumination intensity of the LED array induces long exposure time during image acquisition. To improve the capture efficiency, a variety of techniques have been proposed. In principle, these techniques can be devided into two categories. The first one is to improve the FPM illumination. For example, Kuang \emph{et al.} have been proposed to use a laser instead of the LED array to enhance the intensity of the illumination, so as to reduce the exposure time during the capture procedure~\cite{kuang2015digital}. Other approaches in this category include the lighting up of several LEDs simultaneously~\cite{tian2015computational,dong2014spectral,tian2014multiplexed,zhou2017fourier}, and non-uniform sampling of the object's spectrum~\cite{guo2015optimization,zhang2015self,bian2014content,dong2014sparsely,dong2015resolution} to reduce the required number of images~. These techniques usually change the illumination structure and need to find an optimal strategy, which sacrifices the simplicity of the original FPM. The other one is to improve the reconstruction algorithm~\cite{bian2015fourier,zuo2016adaptive,zhang2017fourier}. For example, the Wirtinger flow optimization algorithm has been proposed to save around 80\% of the exposure time~\cite{bian2015fourier}. But this algorithm increases the calculation cost than the original FPM~\cite{zheng2013wide}. 

In this paper, we propose an alternative approach by analyzing the imaging process. Based on the analysis, we perform simulations and experiments to demonstrate our method. In section 2, we make a theoretical analysis on the FPM imaging process. In section 3 and 4, simulations and experiments are conducted to prove the analysis.

\section{Analysis of the FPM imaging system}
We suppose that, in a FPM system, a complex object with the transmittance function of $O(x,y)=A(x, y)\exp[j\phi(x,y)]$ is located at the front focal plane of the microscopic objective, where $A(x, y)$ and $\phi(x, y)$ are the amplitude and phase of the object. A plane wave parallel to the optical axis illuminates the object with a wavelength of $\lambda$. Without considering the aperture size, the Fourier spectrum $H(u, v)$ located at the back focal plane of the microscopic objective is written as~\cite{goodman2005introduction,Chen_2016_AO}
\begin{align}
H(u, v) = \iint \limits_{-\infty}^{\infty} O(x, y) \exp\left[-j\frac{2\pi}{\lambda f}(xu+yv) \right] \mathrm{d}x\mathrm{d}y, 
\label{eq_H_shift}
\end{align}
where $f$ is the focal length of the objective, and $(u, v)$ is the spatial frequency coordinate. In the case that the illumination angle is $(\theta_x,\theta_y)$, the Fourier spectrum becomes 
\begin{align}
& H_{\theta _x,\theta _y}(u, v) \nonumber\\
= & \iint \limits_{-\infty}^{\infty} O(x, y) \times \exp\left[j2\pi\left(\frac{\sin\theta_x}{\lambda}x + \frac{\sin\theta_y}{\lambda}y \right) \right] \nonumber\\ 
&\times \exp\left[-j\frac{2\pi}{\lambda f}(xu+yv) \right] \mathrm{d}x\mathrm{d}y \nonumber\\
= & H(u-u_0,v-v_0),
\end{align}
which indicates the Fourier spectrum is shifted to $(u_0, v_0)$ due to the illumination angle. It has been proved that $u_0 =f \sin\theta_x$ and $v_0 = f \sin\theta_x$~\cite{teich1991fundamentals}. Because of the limited NA of the system, only a part of the Fourier spectrum transmits the tube lens. Let the radius of the transmitted sub spectrum $H_{\theta _x,\theta _y}^{sub}(u, v)$ be $r$, it then can be written as
\begin{align}
& H_{\theta _x,\theta _y}^{sub}(u, v) \nonumber\\
= & H(u-u_0,v-v_0) \times\text{circ}\left(\frac{\sqrt{u^2+v^2}}{r}\right), 
\end{align}
where $\text{circ}\left({\sqrt{u^2+v^2}}/{r}\right)$ is the circle function. Suppose the tube lens has a same focal length as the objective lens, the captured image under specific illumination angle thus can be written as  
\begin{align}
& I_{\theta _x,\theta _y}(x, y)  \nonumber \\
= &\left|\mathcal{F}^{-1}\left\{ H(u-u_0,v-v_0) \times\text{circ}\left(\frac{\sqrt{u^2+v^2}}{r}\right) \right\} \right|^2  \nonumber\\ 
= &\Bigg| O\left(x,y\right) \times \exp\left[j{2\pi}\left(\frac{\sin\theta_x}{\lambda } x + \frac{\sin\theta_y}{\lambda} y\right) \right] \nonumber \\
& \otimes J\left(x,y\right) \Bigg|^2,   
\label{eq_recon_I}
\end{align}
where $\mathcal{F}^{-1}$ is the inverse Fourier transform, $J(x,y)=\frac{\lambda fr}{\sqrt{x^2+y^2}}J_1\left(\frac{2\pi r\sqrt{x^2+y^2}}{\lambda f}\right) $ is the inverse Fourier transform of the circle function, and $J_1\left( x,y\right) $ is the first order Bessel function.
  
In the case that the object is amplitude-only, i.e., $O(x,y) = A(x,y)$, Eq.~\ref{eq_recon_I} can be written as
\begin{align}
	& I_{\theta _x,\theta _y}(x, y)  \nonumber \\
	= &\Bigg| A(x,y)\times\exp\left[j{2\pi}\left(\frac{\sin\theta_x}{\lambda } x + \frac{\sin\theta_y}{\lambda} y\right) \right] \nonumber \\
	&\otimes J\left(x,y\right) \Bigg|^2.   
	\label{eq_recon_II}
\end{align}
 From Eq.~\ref{eq_recon_II}, it can be seen that the captured intensity images $I_{\theta _x,\theta _y}(x, y)$ equals $I_{-\theta _x,-\theta _y}(x, y)$ strictly.

In the case that the object is phase-only, i.e., $O(x,y) = \exp[j\phi(x,y)]$, Eq.~\ref{eq_recon_I} can be written as
\begin{align}
	& I_{\theta _x,\theta _y}(x, y)  \nonumber \\
	= &\Bigg| \exp\left\lbrace j\left[ \phi(x,y)+ {2\pi}\left(\frac{\sin\theta_x}{\lambda } x + \frac{\sin\theta_y}{\lambda} y\right) \right] \right\rbrace \nonumber \\
	&\otimes J\left(x,y\right) \Bigg|^2.   	
	\label{eq_recon_III}
\end{align}
When the phase-only object is thin enough so that $\left| {2\pi}\left(x \sin\theta_x/\lambda + y \sin\theta_y/\lambda\right)\right| \gg \phi(x,y)$, Eq.~\ref{eq_recon_III} can be written as
 \begin{align}
 & I_{\theta _x,\theta _y}(x, y)  \nonumber \\
 \approx &\left| \exp\left[ j{2\pi}\left(\frac{\sin\theta_x}{\lambda } x + \frac{\sin\theta_y}{\lambda} y\right) \right] \otimes J\left(x,y\right) \right|^2. 
 \label{eq_recon_IV}  	
 \end{align}
From Eq.~\ref{eq_recon_IV}, we can find that the assumption is better satisfied when the $\theta _x$ and $\theta _y$ are larger, which means that the images captured at larger illumination angles are more close to the ground-truth than those that are captured at smaller illumination angles. 
 
In the case that the object is complex, Eq.~\ref{eq_recon_I} can be written as
\begin{align}
	& I_{\theta _x,\theta _y}(x, y)  \nonumber \\
	= &\Bigg| A(x,y)\exp[j\phi(x,y)] \nonumber \\ 
	  &\exp\left[j{2\pi}\left(\frac{\sin\theta_x}{\lambda } x + \frac{\sin\theta_y}{\lambda} y\right) \right] \otimes J\left(x,y\right) \Bigg|^2.   
	\label{eq_recon_V}
\end{align}
Compared to phase-only object, influenced by the amplitude $A(x,y)$, the difference between $I_{\theta _x,\theta _y}(x, y)$ and $I_{-\theta _x,-\theta _y}(x, y)$ is complex, which is hard to compare. 

From the theoretical analysis of the FPM imaging system, we get a preliminary conclusion that the images captured with circular symmetrical illumination angles have little intensity difference when the object is amplitude-only or phase-only. Considering the fact that the FPM uses only intensity images to reconstruct the wide FOV and high resolution images, we speculate that the reconstructed images with a half number of low resolution images can reach a comparable resolution to that with the whole images in the FPM.

\section{Simulation verification}

We performed two simulations to verify the above analysis. In these simulations, we used a 4$\times$ 0.1 NA microscopic objective and a $15 \times 15$ LED array. The wavelength of the LED array was \SI{630}{\nano\meter}, and the spacing between two adjacent LEDs was \SI{4}{\milli\meter}. The distance between the LED array and the sample was \SI{110}{\milli\meter}. The camera sensor has $128\times 128$ pixels, the pitch of which is \SI{6.5}{\micro\meter}.
\begin{figure}[t]
	\centering
	\includegraphics[width=0.3\columnwidth]{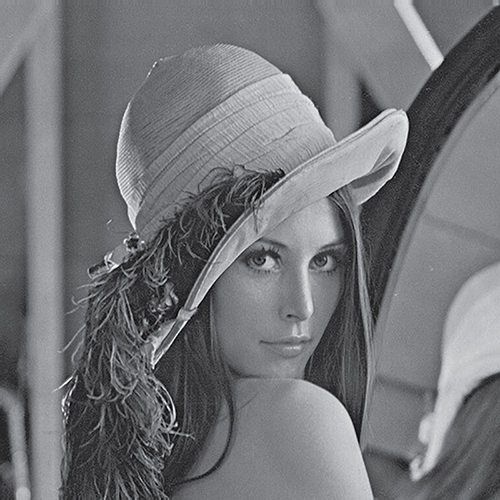}
	\includegraphics[width=0.3\columnwidth]{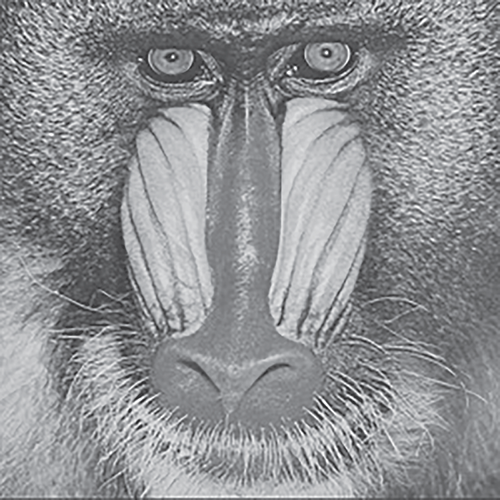}
\caption{\label{fig_sim_obj} Amplitude and phase image of the object used in the simulation.}
\end{figure}

\begin{figure}[t]
	\centering
	\captionsetup[subfigure]{justification=centering}
	\begin{subfigure}[b]{1\columnwidth}
	  	\centering
	  	\includegraphics[width=0.48\columnwidth]{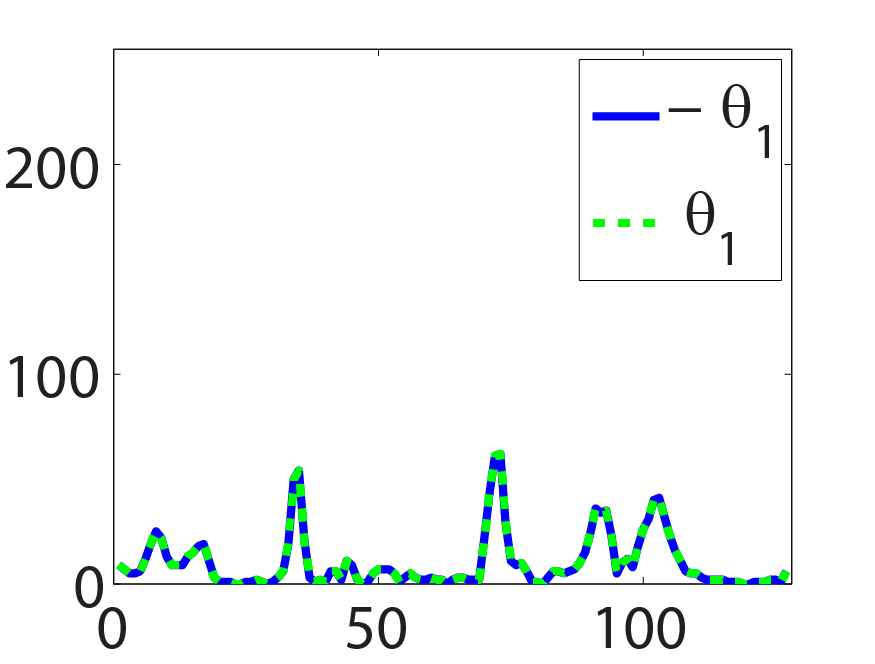}
		\includegraphics[width=0.48\columnwidth]{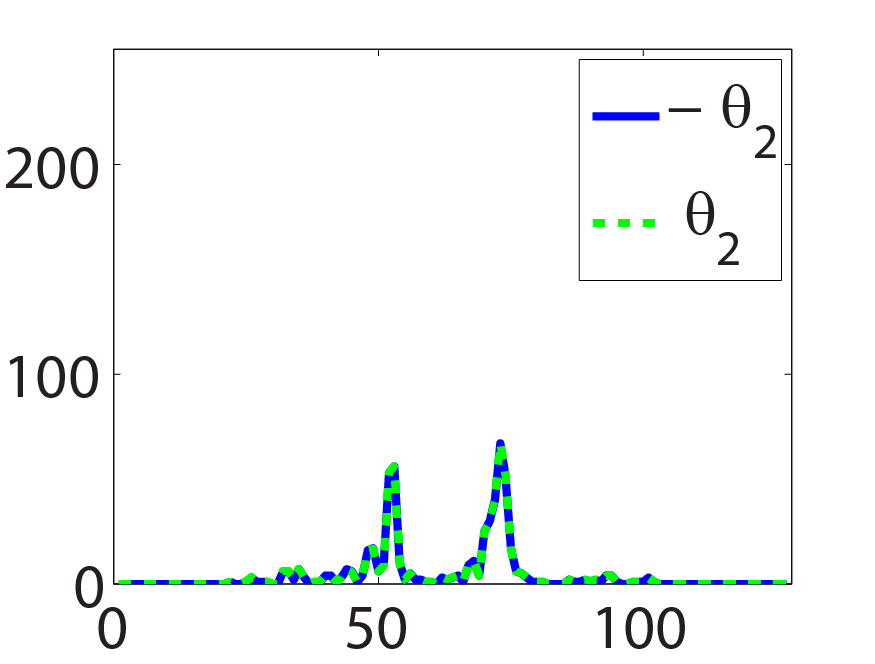}
	    \caption{}
	\end{subfigure}

	\begin{subfigure}[b]{1\columnwidth}
	  	\centering
	  	\includegraphics[width=0.48\columnwidth]{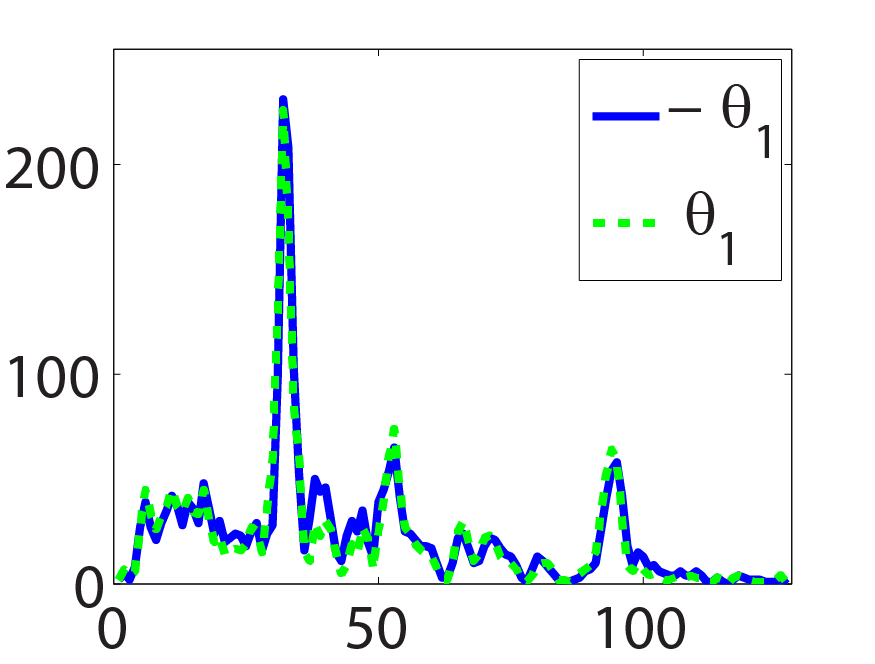}
		\includegraphics[width=0.48\columnwidth]{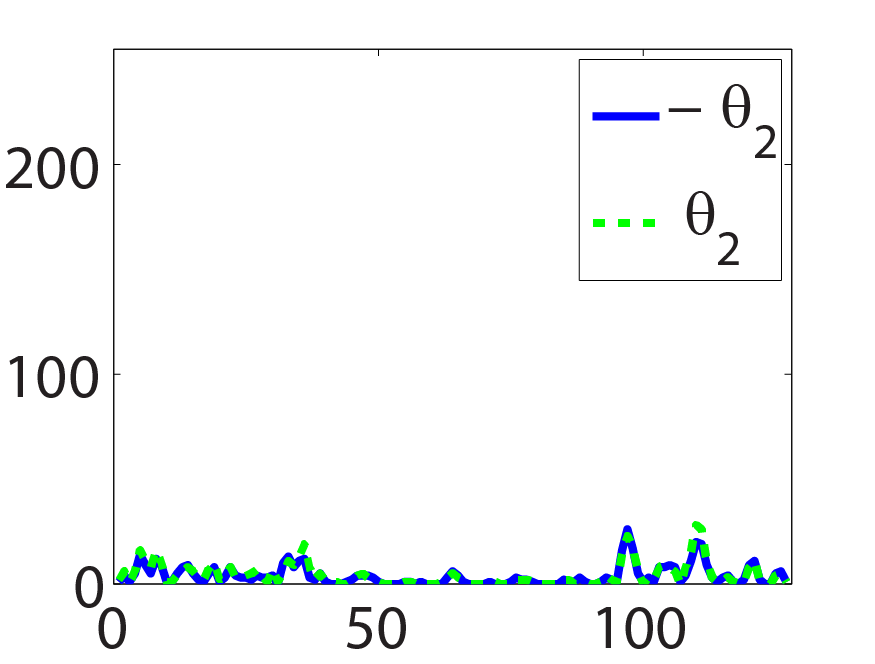}
	    \caption{}
	\end{subfigure}

	\begin{subfigure}[b]{1\columnwidth}
	  	\centering
	  	\includegraphics[width=0.48\columnwidth]{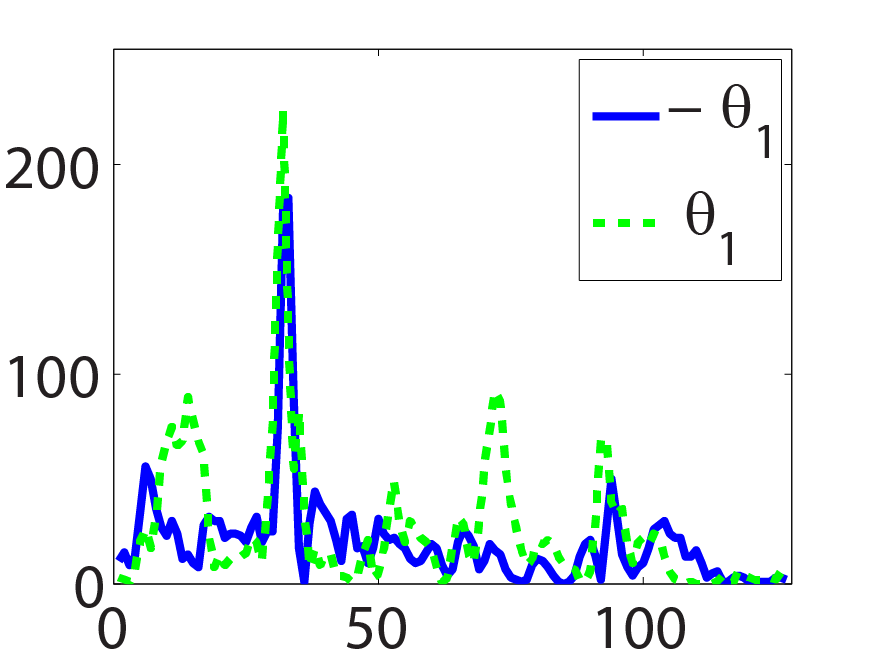}
		\includegraphics[width=0.48\columnwidth]{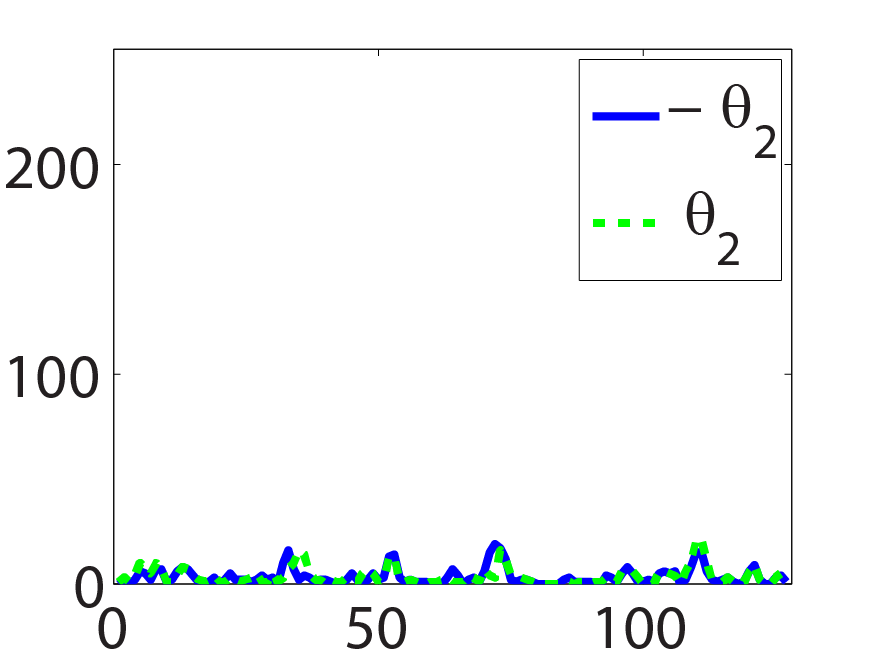}
	    \caption{}
	\end{subfigure}
\caption{\label{fig_sim_capture1} Simulation 1: Comparison of the images captured under symmetrical illumination angles using (a) amplitude-only, (b) phase-only and (c) complex objects.}
\end{figure}

In the first simulation, we compared the images captured by a microscope under symmetrical illumination angles. Figure~\ref{fig_sim_obj} shows the amplitude and phase profiles of the object used in this simulation. The original object has $512 \times 512$ pixels. The pixel number of the low resolution pictures captured by the camera is $128 \times 128$. To make the object thin enough, the range of phase value was set to $[0, 0.5 \pi]$. The object was set to be at the focal plane of the microscope. 

To compare the captured images, we plot the gray values along the horizontal center lines of the captured images and calculate the root-mean-square error~(RMSE) between them. The images were compared in three cases, i.e., the object is amplitude-only, phase-only and complex. Figure~\ref{fig_sim_capture1} shows the results. $\theta_1$ and $\theta_2$ are the illumination angles. $\theta_1$ is $(\theta_x, \theta_y) = (4.2^\circ, 4.2^\circ)$ and $\theta_2$ is $(\theta_x,\theta_y)=(8.3^\circ, 8.3^\circ)$. Figure~\ref{fig_sim_capture1}(a) shows the results when the object is amplitude-only. The RMSE of the images are all 0 in $\theta_1$ and $\theta_2$ respectively, we can observe that the curves are exactly identical.  Figure~\ref{fig_sim_capture1}(b) shows the results when the object is phase-only. The RMSE of the images are 5.71 and 3.63 in $\theta_1$ and $\theta_2$ respectively, we can observe that the difference between the curves become smaller when the illumination angle $\theta_1$ is increasing. 
Figure~\ref{fig_sim_capture1}(c) shows the results of the complex object. The RMSE of the images are 28.01 and 5.65 in $\theta_1$ and $\theta_2$ respectively. There is obvious difference between the curves in different illumination angles. 
The simulation results are coincided with the analysis in section 2. Considering both the numerical analysis and the simulations, we come to a conclusion that the captured images under symmetrical illumination angles are absolutely the same when the sample is amplitude-only, are almost the same when the sample is thin phase-only, and are different when the object is complex.
\begin{figure}[!t]
	\centering
	\captionsetup[subfigure]{justification=centering}

	\begin{subfigure}[b]{1\columnwidth}		
		\includegraphics[width=0.28\columnwidth]{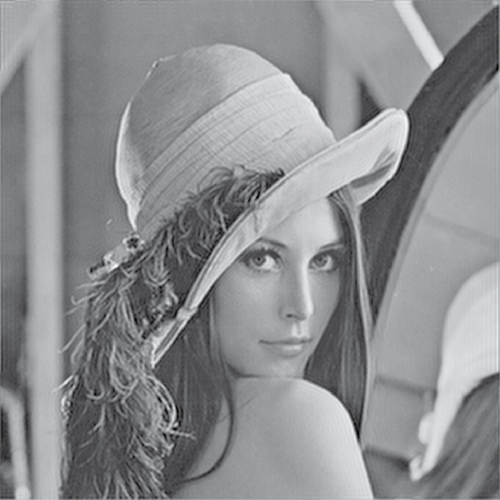}
		\includegraphics[width=0.28\columnwidth]{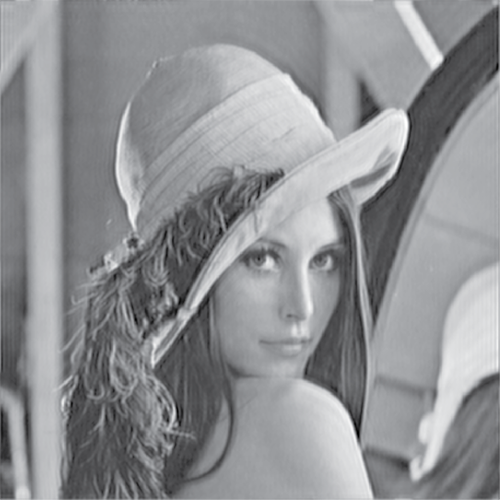}
		\includegraphics[width=0.36\columnwidth]{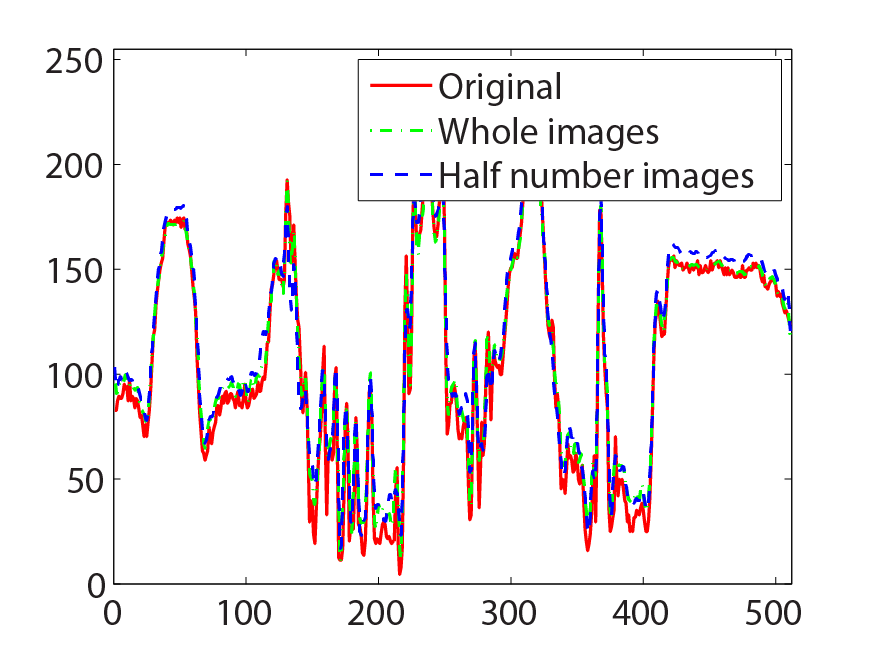}
		\caption{Amplitude-only object reconstructions.}
	\end{subfigure}

	\begin{subfigure}[b]{1\columnwidth}
		\includegraphics[width=0.28\columnwidth]{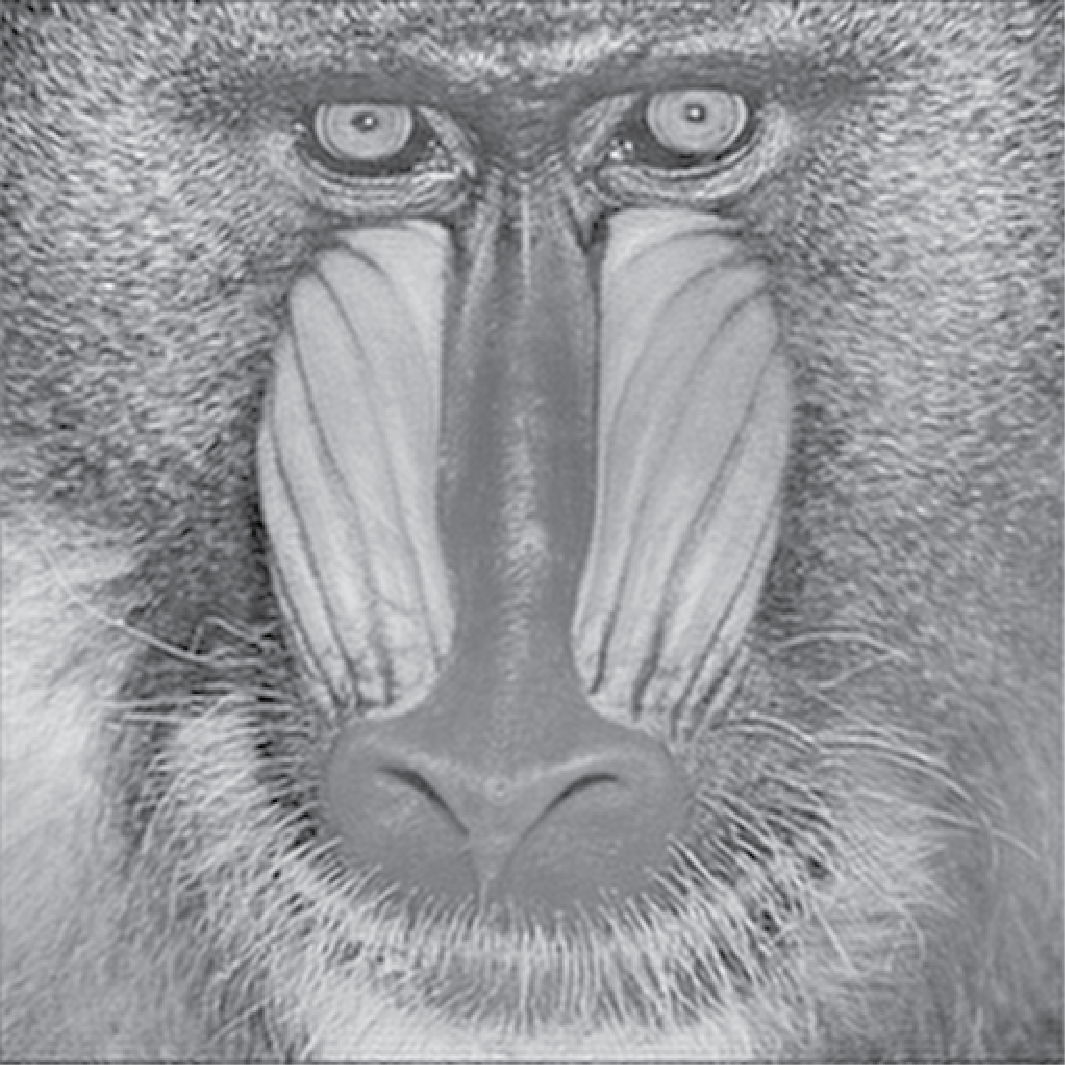}
		\includegraphics[width=0.28\columnwidth]{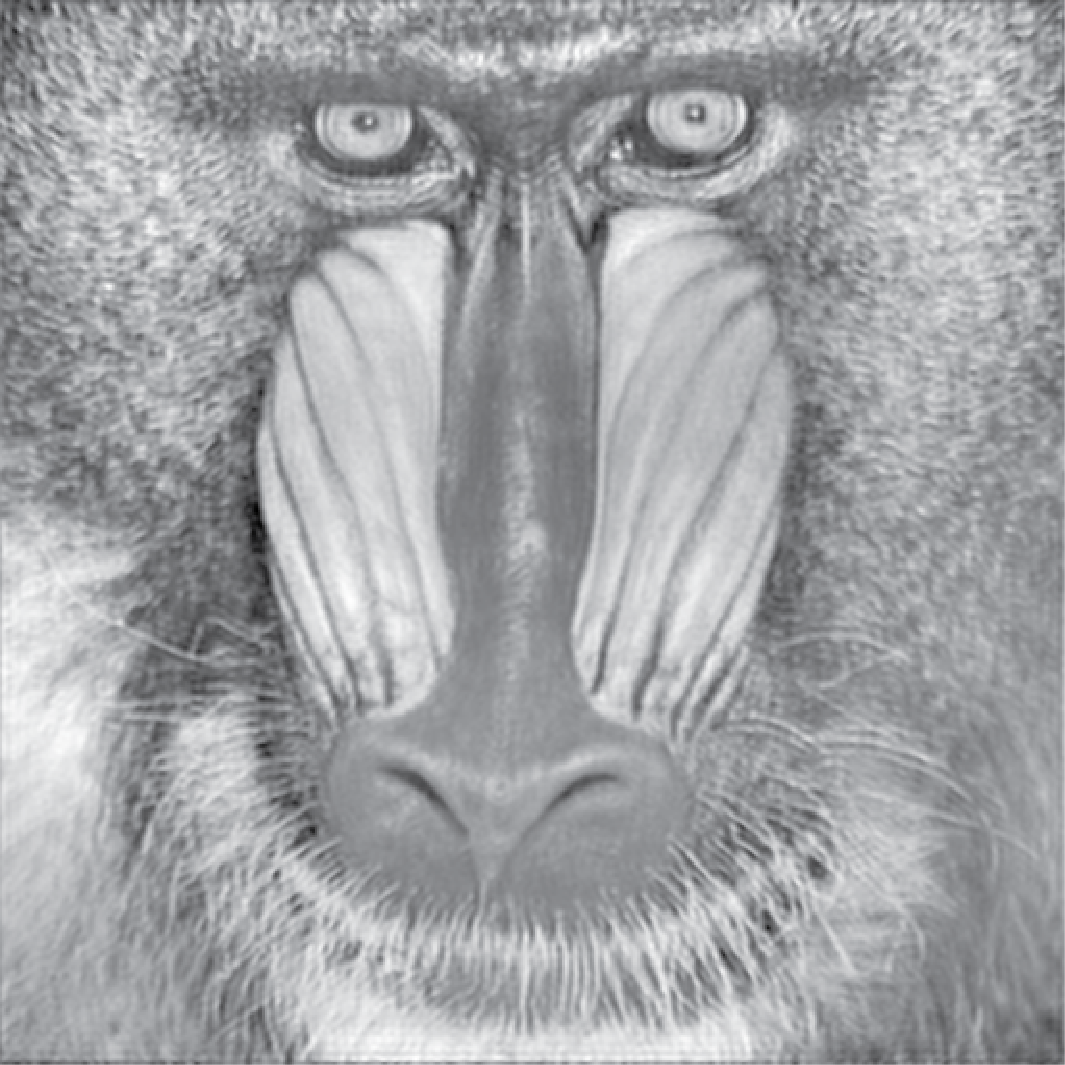}
		\includegraphics[width=0.36\columnwidth]{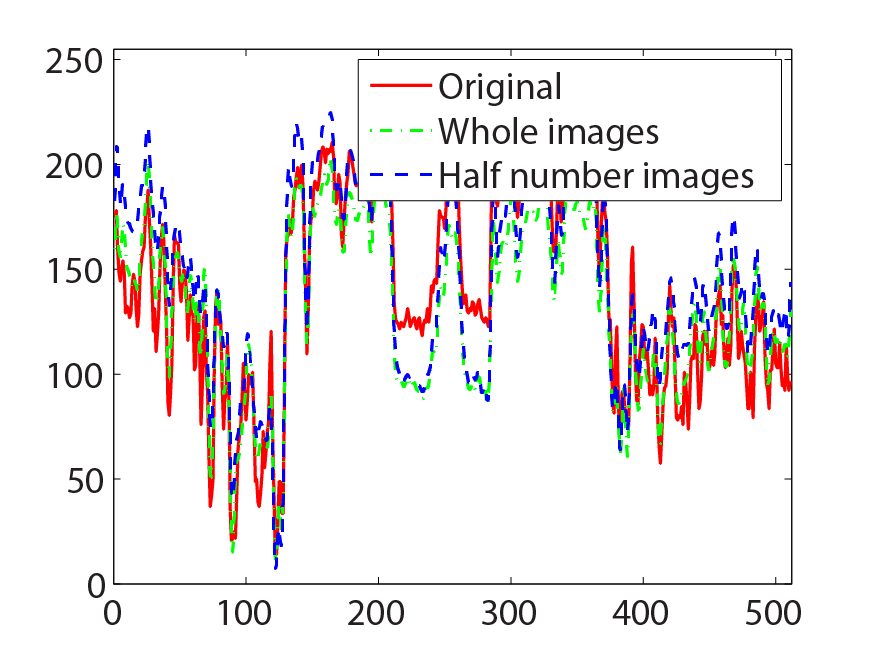}
		\caption{Phase-only object reconstructions.}
	\end{subfigure}

	\begin{subfigure}[b]{1\columnwidth}
		\includegraphics[width=0.28\columnwidth]{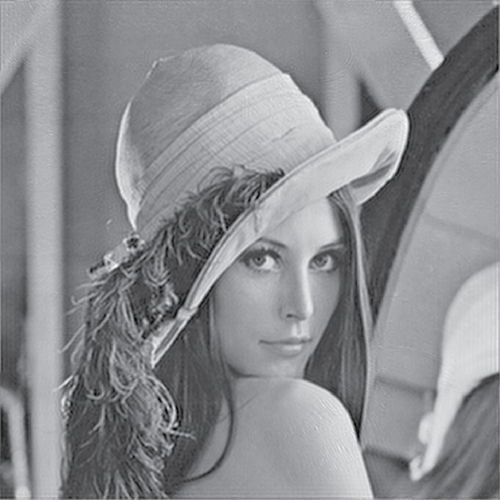}
		\includegraphics[width=0.28\columnwidth]{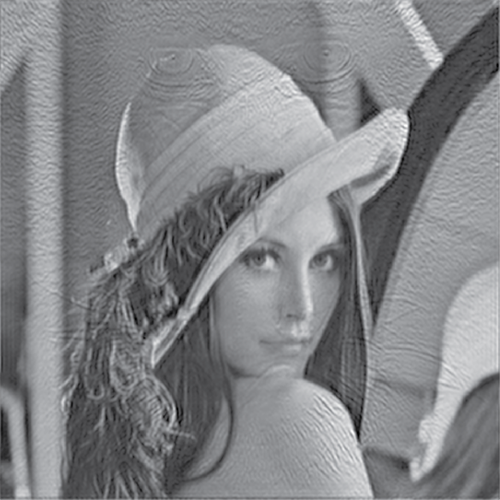}
		\includegraphics[width=0.36\columnwidth]{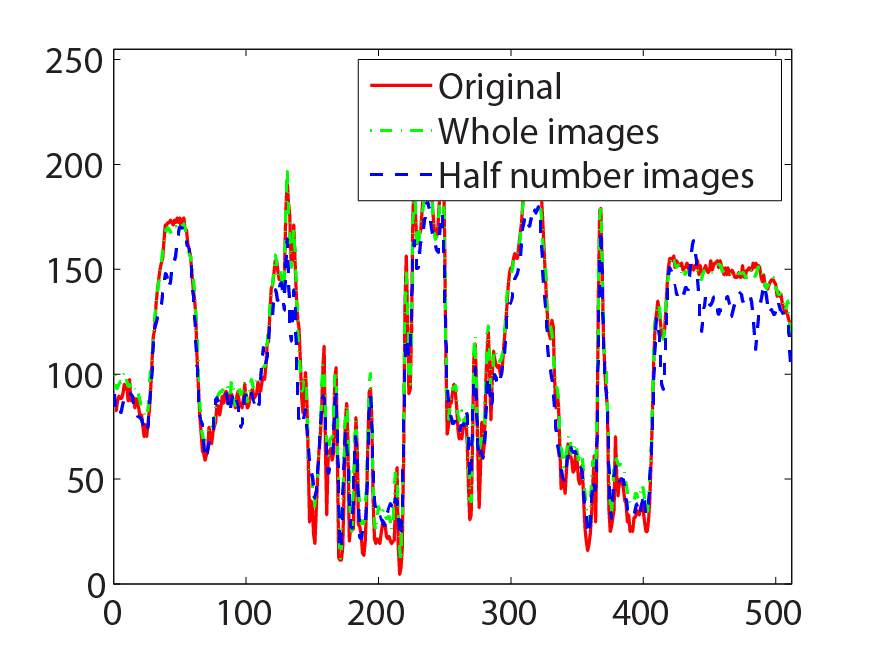}
		\vspace{1pt}

		\includegraphics[width=0.28\columnwidth]{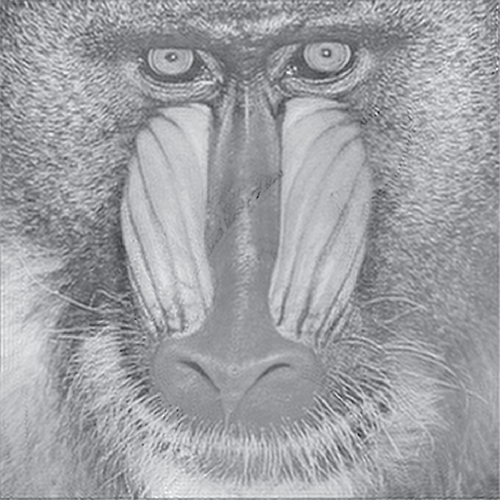}
		\includegraphics[width=0.28\columnwidth]{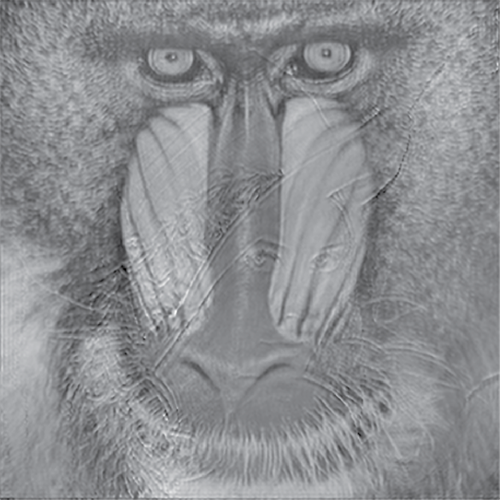}
		\includegraphics[width=0.36\columnwidth]{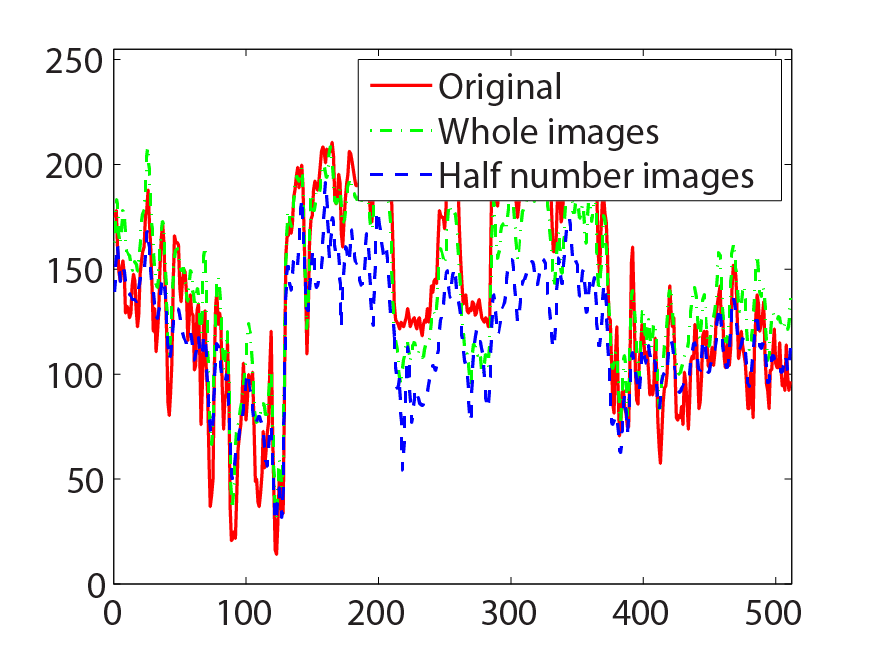}

		\caption{Complex object reconstruction.}
	\end{subfigure}
\caption{\label{fig_sim_fpm} Simulation 2: Comparison of the FPM reconstructions using a/an (a) phase-only, (b) amplitude-only, and (c) complex object. Compared with the images in the left column, the RMSE of the images in the center column are 6.25, 17.23, 12.82, 28.84 respectively.}
\end{figure}

The second simulation was performed to verify our speculation. The amplitude and phase profiles of the object used in the FPM simulation are the same as in the first simulation. The results are plotted in Fig.~\ref{fig_sim_fpm}. 
In Fig.~\ref{fig_sim_fpm}(a-c), the left and center columns show the reconstructed amplitude and phase with whole $(15\times15)$ and half $(15\times8)$ camera images respectively. The right column shows the intensity comparison between the left and center columns along the middle horizontal intersection. Figure~\ref{fig_sim_fpm}(a) and (b) show that the shape and distribution of the amplitude and phase profiles are similar. We also performed a simulation when the object is complex. From the results shown in Fig.~\ref{fig_sim_fpm}(c), we can observe obvious difference between using whole and half captured images. The reason is that the captured images are totally different when the object is illuminated by symmetrical angles. Besides, during the iteration process of phase retrieval algorithm in FPM, there are some crosstalk between the amplitude and phase reconstructions. From the simulation results, we observe that the reconstructions using half captured images do not show obvious resolution degradation in the cases of amplitude-only and phase-only objects. 

\section{Experimental verification}
We built an FPM system by replacing the light source of an Olympus IX 73 inverted microscope with a programmable LED array~($32 \times 32$ LEDs, \SI{4}{\milli\meter} spacing) controlled by an Arduino. The LEDs have a central wavelength of \SI{629}{\nano\meter} and a bandwidth of \SI{20}{\nano\meter}. 
 
Two experiments were performed to verify our speculation. One used a USAF-1951 resolution target as the object, which can be regarded as an amplitude-only object. The other one used a microscopic object that has phase information. Both samples were imaged with a 4$\times$ 0.1 NA microscopic objective and a complementary metal oxide semiconductor~(CMOS) camera~(PCO. edge 4.2). The pixel size of the CMOS is \SI{6.5}{\micro\meter}. 

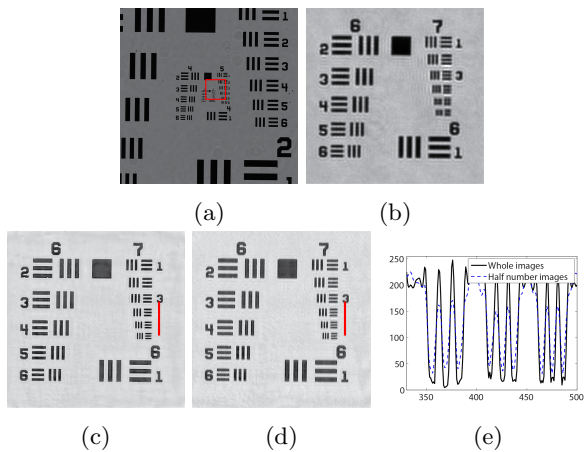
\begin{figure} [ht]
\centering
	\begin{subfigure}[b]{0.3\columnwidth}
	\centering
		\begin{tikzpicture}[scale=1, transform shape, font=\Huge]
		    \scope[nodes={inner sep=0,outer sep=0}]
		    \node[anchor=south west] 
		    {
		     \includegraphics[width=1\columnwidth]{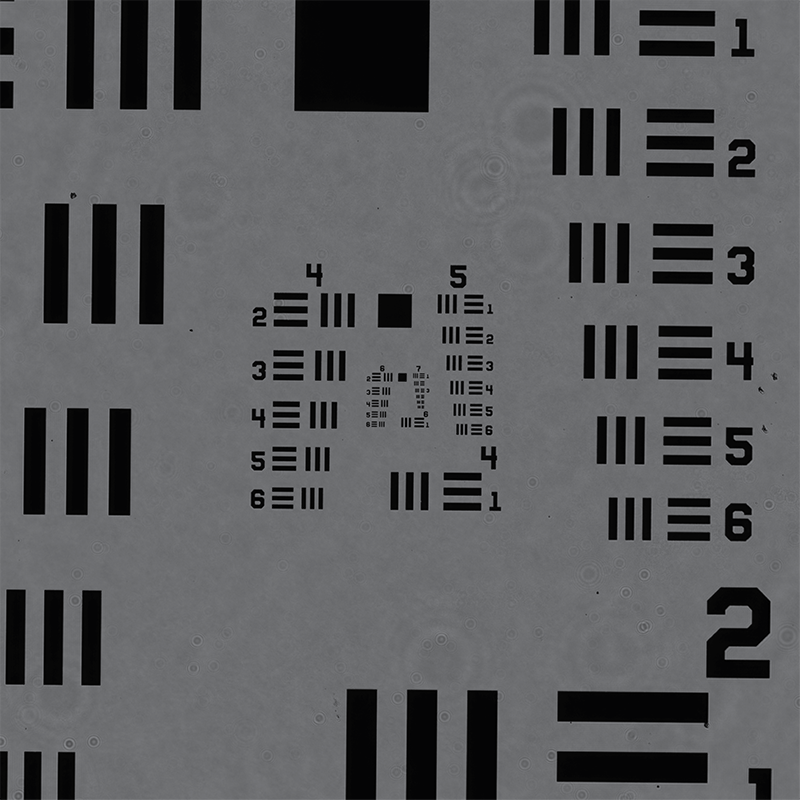}

	        };
		    \draw [red] (1.12, 1.12) rectangle(1.38, 1.38);
		    \endscope
		\end{tikzpicture}
	    \caption{}
	\end{subfigure}
	\begin{subfigure}[b]{0.3\columnwidth}
		\includegraphics[width=1\columnwidth]{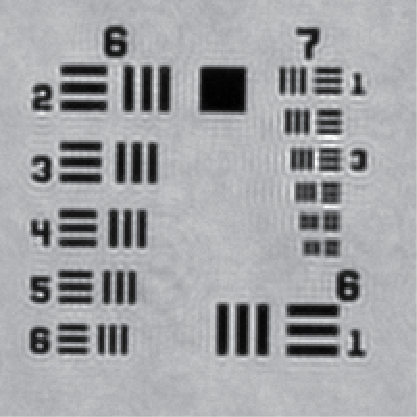}
	    \caption{}
	\end{subfigure}
	
	\begin{subfigure}[b]{0.3\columnwidth}
		\centering
		\begin{tikzpicture}[scale=1, transform shape, font=\Huge]
		    \scope[nodes={inner sep=0,outer sep=0}]
		    \node[anchor=south west] 
		    {
		     \includegraphics[width=1\columnwidth]{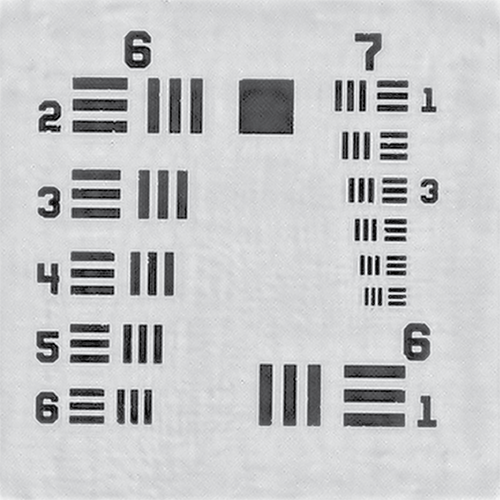}
	        };
		    \draw [red,thick] (2, 0.95)--(2, 1.4) ;
		    \endscope
		\end{tikzpicture}
	    \caption{}
	\end{subfigure}
	\begin{subfigure}[b]{0.3\linewidth}
		\centering
		\begin{tikzpicture}[scale=1, transform shape, font=\Huge]
		    \scope[nodes={inner sep=0,outer sep=0}]
		    \node[anchor=south west] 
		    {
		     \includegraphics[width=1\columnwidth]{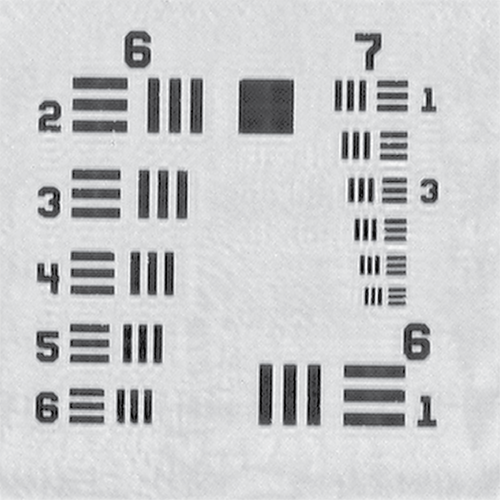}

	        };
		    \draw [red,thick] (2, 0.95)--(2, 1.4) ;
		    \endscope
		\end{tikzpicture}
	    \caption{}
	\end{subfigure}
	    \begin{subfigure}[b]{0.37\linewidth}
		\centering
		\includegraphics[width=1\linewidth]{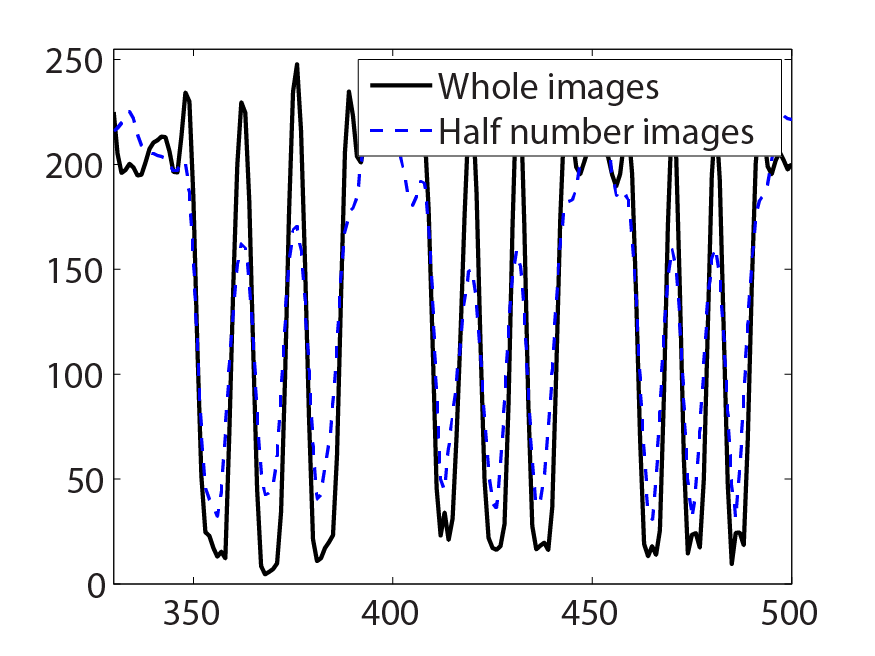}
	    \caption{}
	\end{subfigure}
	\caption{\label{fig_exp_usaf} (a) is the captured raw image of the USAF resolution target, (b) is image within the red rectangle in (a). (c)(d) are the reconstructed amplitudes with the whole and half of the captured images. Compared with (c), the RMSE of (d) is 10.90. (e) is the amplitude profiles along the red horizontal lines in (c) and (d). }
\end{figure} 

In the first experiment, the central $15 \times 15$ LEDs of the array were on sequentially so that 225 low resolution images were captured by the sCMOS camera.  The exposure time of the camera was \SI{600}{\milli\second} per acquisition. The distance between the LED array and the sample was \SI{108}{\milli\meter}. The expected synthesized NA of the imaging system is 0.45. The experimental results are shown in Fig.~\ref{fig_exp_usaf}. Figure~\ref{fig_exp_usaf}(b) is one of the segment of the original captured images. Figure~\ref{fig_exp_usaf}(c) shows the reconstructed result using 225 images. Figure~\ref{fig_exp_usaf}(d) shows the reconstructed result using 120 images. Figure~\ref{fig_exp_usaf}(e) shows the intensity profile along the red vertical red lines in Figs.~\ref{fig_exp_usaf}(c) and \ref{fig_exp_usaf}(d). From the two curves, we can observe that their shape are almost the same, but the gray values of the reconstruction with all the acquired images are higher than that using half of them. This means that the image resolution of Figs.~\ref{fig_exp_usaf}(c) and \ref{fig_exp_usaf}(d) has no significant difference except the image contrast.

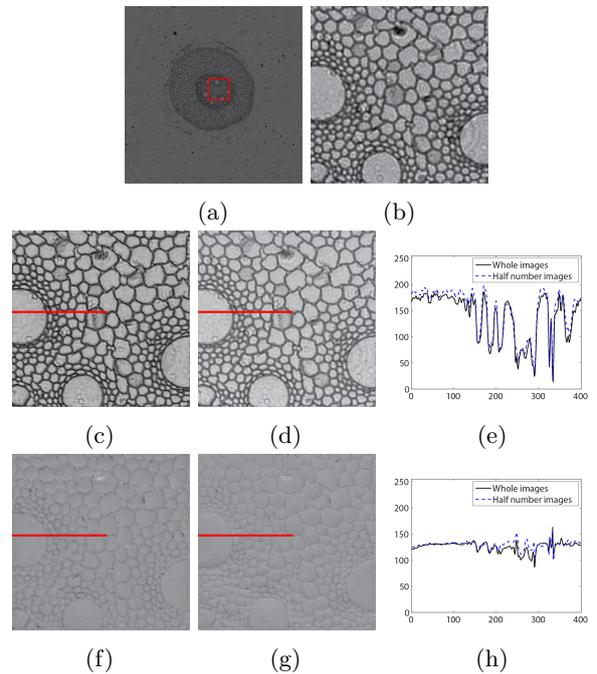
\begin{figure} [ht]
 \centering
	\begin{subfigure}[b]{0.3\columnwidth}
		\begin{tikzpicture}[scale=1, transform shape, font=\Huge]
		    \scope[nodes={inner sep=0,outer sep=0}]
		    \node[anchor=south west] 
		    {
		     \includegraphics[width=1\columnwidth]{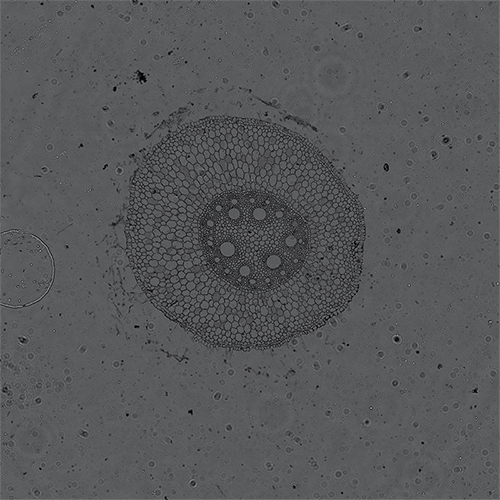}

	        };
		    \draw [red] (1.1, 1.1) rectangle(1.38, 1.38);
		    \endscope
		\end{tikzpicture}
	    \caption{}
	\end{subfigure}
	\begin{subfigure}[b]{0.3\columnwidth}
		\includegraphics[width=1\columnwidth]{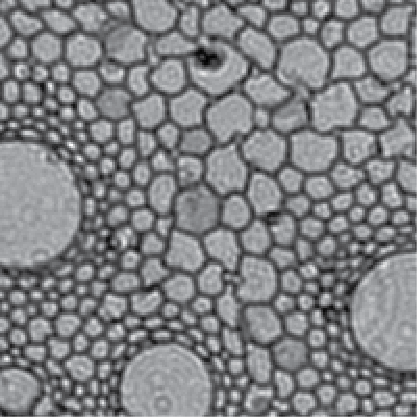}
	    \caption{}
	\end{subfigure}
	
	\begin{subfigure}[b]{0.3\linewidth}
		\centering
		\begin{tikzpicture}[scale=1, transform shape, font=\Huge]
		    \scope[nodes={inner sep=0,outer sep=0}]
		    \node[anchor=south west] 
		    {
		     \includegraphics[width=1\columnwidth]{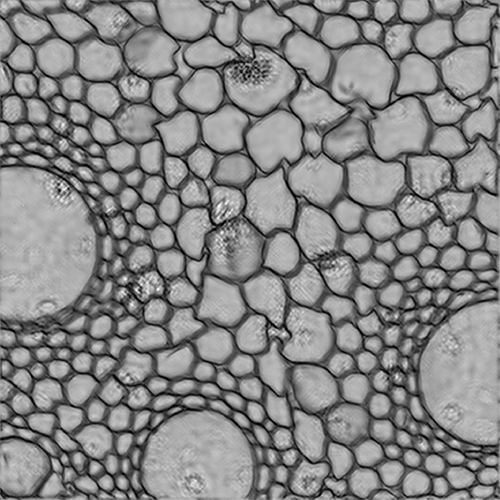}

	        };
		    \draw [red,thick] (0, 1.25)--(1.25, 1.25) ;
		    \endscope
		\end{tikzpicture}
	    \caption{}
	\end{subfigure}
	    \begin{subfigure}[b]{0.3\linewidth}
		\centering
		\begin{tikzpicture}[scale=1, transform shape, font=\Huge]
		    \scope[nodes={inner sep=0,outer sep=0}]
		    \node[anchor=south west] 
		    {
		     \includegraphics[width=1\columnwidth]{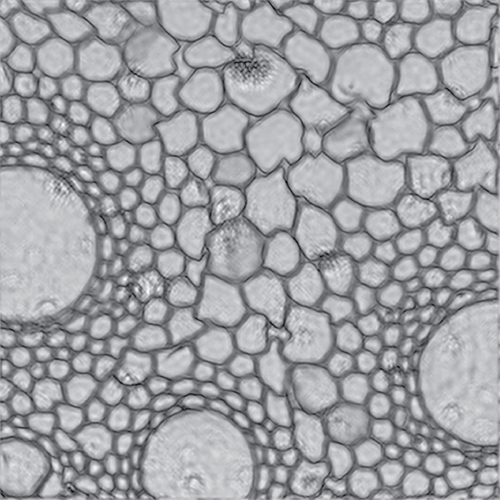}

	        };
		    \draw [red,thick] (0, 1.25)--(1.25, 1.25) ;
		    \endscope
		\end{tikzpicture}
	    \caption{}
	\end{subfigure}
	    \begin{subfigure}[b]{0.37\linewidth}
		\centering
		\includegraphics[width=1\linewidth]{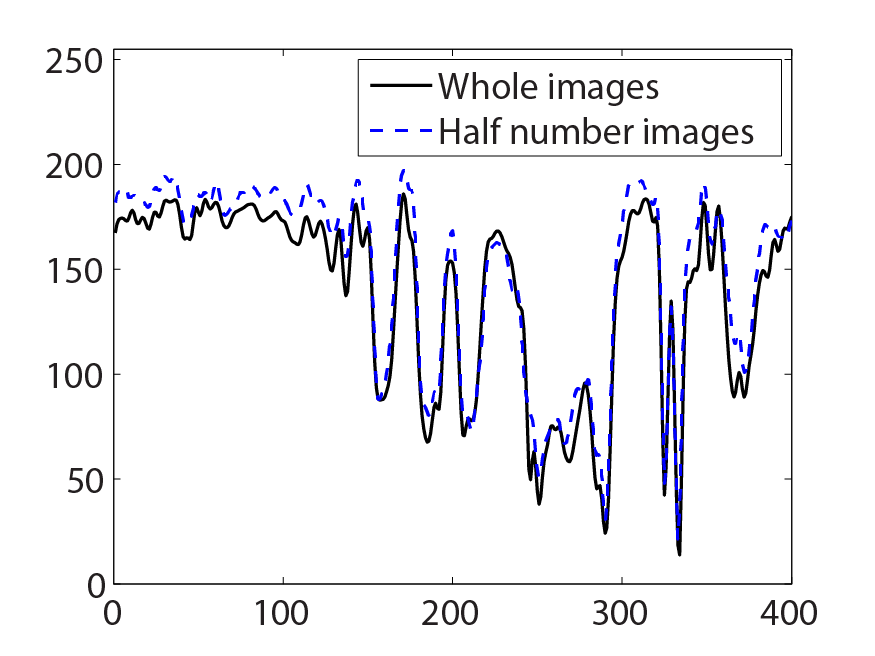}
	    \caption{}
	\end{subfigure}

	\begin{subfigure}[b]{0.3\linewidth}
		\centering
		\begin{tikzpicture}[scale=1, transform shape, font=\Huge]
		    \scope[nodes={inner sep=0,outer sep=0}]
		    \node[anchor=south west] 
		    {
		     \includegraphics[width=1\columnwidth]{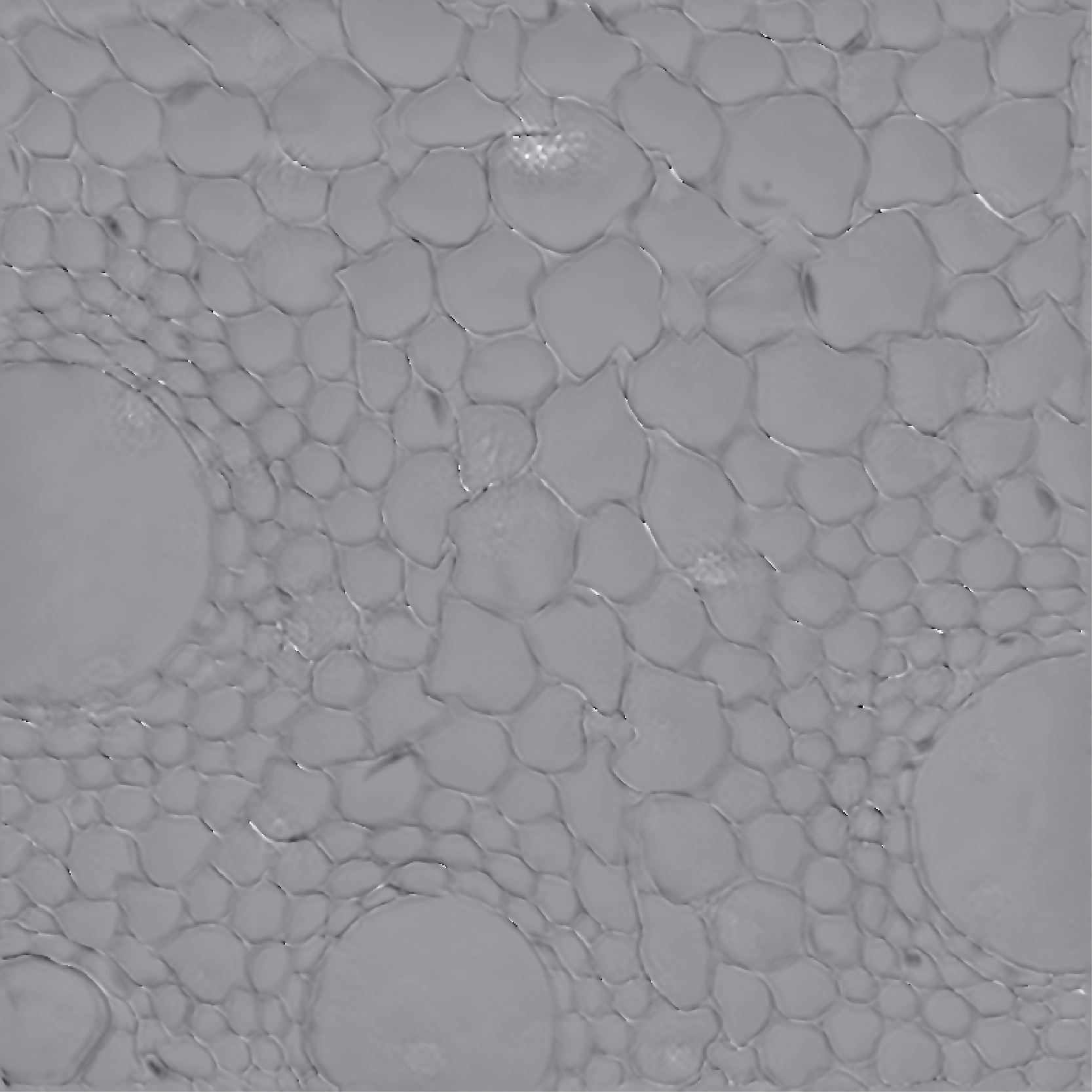}

	        };
		    \draw [red,thick] (0, 1.25)--(1.25, 1.25) ;
		    \endscope
		\end{tikzpicture}
	    \caption{}
	\end{subfigure}
	    \begin{subfigure}[b]{0.3\linewidth}
		\centering
		\begin{tikzpicture}[scale=1, transform shape, font=\Huge]
		    \scope[nodes={inner sep=0,outer sep=0}]
		    \node[anchor=south west] 
		    {
		     \includegraphics[width=1\columnwidth]{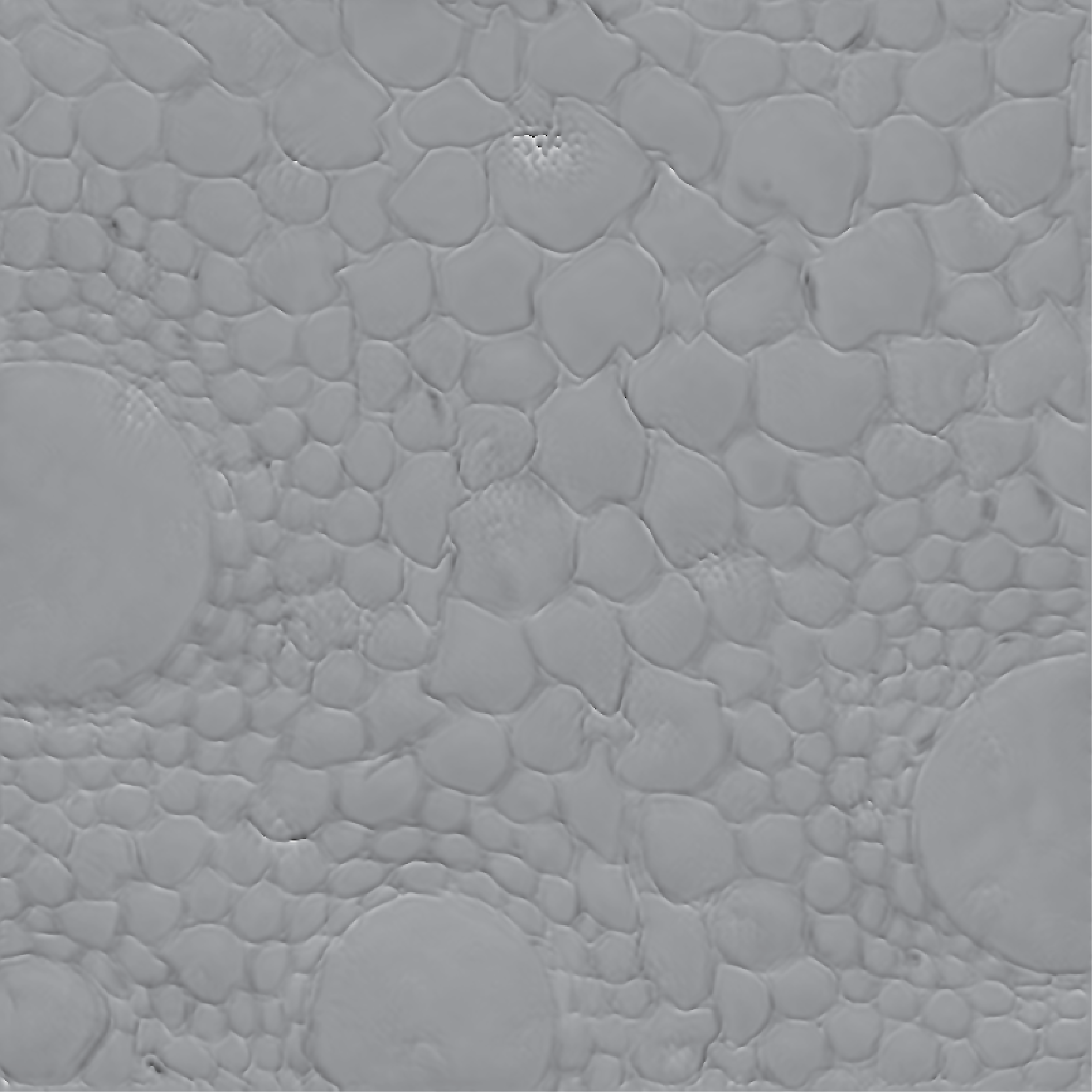}

	        };
		    \draw [red,thick] (0, 1.25)--(1.25, 1.25) ;
		    \endscope
		\end{tikzpicture}
	    \caption{}
	\end{subfigure}
	    \begin{subfigure}[b]{0.37\linewidth}
		\centering
		\includegraphics[width=1\columnwidth]{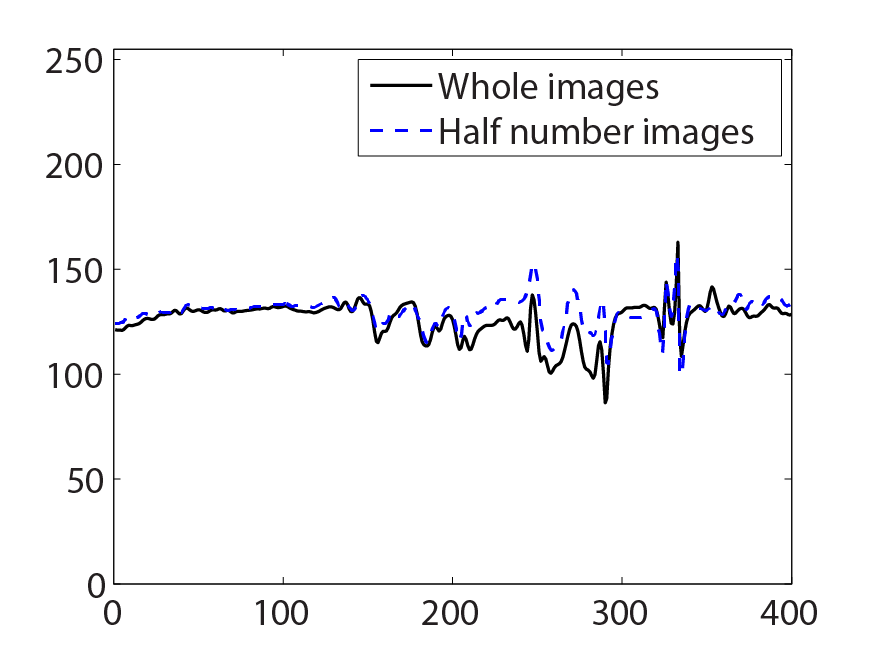}
	    \caption{}
	\end{subfigure}
	\caption{\label{fig_exp_micro}(a) is the captured raw image of a plant root, (b) is the image within the red rectangle in (a). (c)(d) are the reconstructed amplitudes with the whole and half of the captured images. Compared with (c),The RMSE of (d) is 12.91. (f)(g) are the reconstructed phases with the whole and half of the captured images. Compared with (f), the RMSE of (g) is 10.67. (e) is the amplitude profiles along the red horizontal lines in (c) and (d). (h) is the phase profiles along the red horizontal lines in (f) and (g).}
\end{figure} 

In the second experiment, the central $17\times17$ LEDs of the array were on sequentially so that 289 low resolution images were captured. The distance between the LED array and the sample was \SI{113.5}{\milli\meter}. The expected synthesized NA of the imaging system is 0.48. The other experimental specifications were the same as in the first experiment. The experimental results are shown in Fig.~\ref{fig_exp_micro}. Figure~\ref{fig_exp_micro}(b) is one of the segment of the original captured images.  Figure~\ref{fig_exp_micro}(c) and \ref{fig_exp_micro}(f) show the reconstructed amplitudes and phases using 289 images. Figure~\ref{fig_exp_micro}(d) and \ref{fig_exp_micro}(g) show the reconstructed amplitude and phase using 153 images. Figure\ref{fig_exp_micro}(e) shows the intensity profile along the red horizon lines in Fig.~\ref{fig_exp_micro}(c) and \ref{fig_exp_micro}(d). Figure~\ref{fig_exp_micro}(h) shows the intensity profile along the red horizon lines in Figs.~\ref{fig_exp_micro}(f) and \ref{fig_exp_micro}(g). We can also observe that the reconstructions don't show obvious resolution degradation both in the reconstructed amplitudes and phases, only the image contrast reduction can be observed. Because the real object's amplitude and phase exist similarity in structure, they perform like a phase-only object. Therefor, there is no significant crosstalk between the amplitude and phase.

\section{Conclusion}
Under the assumption of thin biological samples of FPM, both the theoretical analysis and simulations of a microscopic imaging system have shown that images captured with circular symmetrical illumination angles have little intensity difference when the sample is amplitude-only or phase-only, therefore, the FPM reconstructions show insignificant difference. In the FPM experiments, both amplitude-only and complex objects were used to perform the verification. The reconstructions didn't show obvious resolution degradation between using full and half images, only the image contrast reduction can be observed. It also shows the reconstructions are degraded very less when the phase and amplitude of the object have similar distribution. Since many biological samples satisfy this condition, half number of captured images can be used in these cases. Owing to half number images reduction, time costs of both image capture and computational processing were also reduced in half.

\section*{Acknowledgments}  
This work was supported by the National Natural Science Foundation of China~(61705241, 61327902 and 61377005), Natural Science Foundation of Shanghai~(17ZR1433800), and Chinese Academy of Sciences~(QYZDB-SSW-JSC002). 


\end{document}